\documentstyle[aps,multicol,eqsecnum,epsfig]{revtex}
\def\be{\begin{eqnarray}}
\def\ee{\end{eqnarray}}
\def\l{\langle}
\def\r{\rangle}

\begin{document}
\title{Optimal Manipulations with Qubits: 
Universal Quantum Entanglers
}
\author{
Vladim\'{\i}r Bu\v{z}ek$^{1,2}$ and Mark Hillery$^{3}$
}
\address{
$^{1}$
Institute of Physics, Slovak Academy of Sciences, D\'ubravsk\'a cesta 9,
842 28 Bratislava, Slovakia\newline
$^{2}$
Faculty of Informatics, Masaryk University, Botanick\'a 68a, Brno 602 00,
Czech Republic\newline
$^{3}$
Department
of Physics and Astronomy, Hunter College, CUNY,
695 Park Avenue, New York, NY 10021, USA\newline
}

\date{21 March 2000}
\maketitle
\begin{abstract}
We analyze various scenarios for entangling 
two initially unentangled  qubits. In particular,  
we propose an optimal universal entangler which entangles a qubit
in unknown state $|\Psi\rangle$ with a qubit in a reference
(known) state $|0\rangle$.
 That is, our entangler generates the output state
which is as close as possible to the pure 
(symmetrized) state $(|\Psi\rangle|0\rangle +|0\rangle|\Psi\rangle)$.
The most attractive feature of this entangling
machine, is that the fidelity of its performance
(i.e. the distance between the output and the ideally entangled --
symmetrized  state)
 does not depend
on the input  and takes the constant value
${\cal F}= (9+3\sqrt{2})/14\simeq 0.946$.
We also analyze  how to optimally generate
from a single qubit initially prepared in an unknown state $|\Psi\r$
a two qubit entangled system which is as close as possible to
a Bell state $(|\Psi\r|\Psi^\perp\r+|\Psi^\perp\r|\Psi\r)$,
where $\l\Psi|\Psi^\perp\r =0$. 
\end{abstract}

\vspace{-0.3cm}
\begin{multicols}{2}

%\vspace{-0.5cm}

\section{Introduction}
\label{sec1}
A pure quantum state of two systems $A$ and $B$ is said to be
entangled if it is not a product of a state for $A$ and a state
for $B$.  Two systems in an entangled state are correlated,
and these correlations are intrinsically quantum mechanical
\cite{Peres1}.
For example, one must use entangled states in order to
produce violations of Bell inequalities or in the test of
local realism proposed by Hardy \cite{Hardy1,Hardy2}.  
Entangled states
also play a key role in quantum information, in particular
they are essential in quantum teleportation \cite{Bennett1}
and in superdense coding \cite{Bennett2}.  
In quantum computers entanglement is
one of the features of quantum mechanics which give these 
machines their power \cite{Steane1}.

Here we would like to consider the problem of
how to produce entanglement.  In
particular, if we are given particles, or systems, $A$ and $B$ in 
the pure states $|\Psi\rangle_{_A}$ and $|\Phi\rangle_{_B}$, 
respectively, we would like to produce the state 
$(|\Psi\rangle_{_A}
|\Phi\rangle_{_B} +|\Phi\rangle_{_A}|\Psi\rangle_{_B})$ (up to
normalization).
Formally, we are looking for the symmetrization map
\be
{\cal S}: |\Psi\rangle|\Phi\rangle \longrightarrow 
(|\Psi\rangle
|\Phi\rangle +|\Phi\rangle|\Psi\rangle).
\label{1.1}
\ee
In what follows, where possible 
 we omit explicit subscripts $A$ and $B$. The order in which
the vectors are written in the tensor products implicitly denotes
to which system they belong
(i.e. the left vector corresponds to the system $A$, while
the right vector corresponds to the system $B$).
We assume that the two quantum systems (e.g., qubits) are 
physically distinguishable. For instance they could be
located in different regions of space. The task is to entangle 
their internal degrees of freedom.

That the symmetrization cannot be done perfectly via a unitary
transformation  can be 
shown by the following argument. We consider the case
in which $|\Psi\rangle$ and $|\Phi\rangle$ are both qubits.
A perfect
transformation would have to transform the basis vectors as
\begin{eqnarray}
|00\rangle |v_{0}\rangle &\rightarrow &|00\rangle |v_{1}\rangle 
\nonumber \\
|01\rangle |v_{0}\rangle &\rightarrow &\frac{1}{\sqrt{2}}
(|01\rangle +|10\rangle )|v_{2}\rangle \nonumber \\
|10\rangle |v_{0}\rangle &\rightarrow &\frac{1}{\sqrt{2}}
(|01\rangle +|10\rangle )|v_{3}\rangle \nonumber \\
|11\rangle |v_{0}\rangle &\rightarrow & |11\rangle |v_{4}\rangle ,
\label{1.2}
\end{eqnarray}
where the $|v_{j}\rangle$, for $j=0,4$, are normalized 
``machine'' vectors, i.\ e.\ we assume that the entangler itself
has its own degrees of freedom. In addition, it is assumed 
that the entangler is always initially in the same state,
$|v_{0}\rangle$.  Unitarity requires that 
$\langle v_{2}|v_{3}\rangle =0$.  Now
let us consider the case where the input vectors are 
$|\Psi\rangle=\alpha|0\r+\beta|1\r$
 and $|\Phi\r =|0\rangle$ (i.e. the state of the qubit $A$ is unknown, while
the qubit $B$ is in a known state).
  The  transformation (\ref{1.2}) gives us
\begin{equation}
|\Psi\rangle |0\rangle \rightarrow 
\alpha |00\rangle 
|v_{1}\rangle +\frac{\beta}{\sqrt{2}}(|01\rangle +
|10\rangle )|v_{3}\rangle ,
\label{1.3}
\end{equation}
whereas what it should produce is a vector proportional
to $|\Psi\rangle|0\rangle +|0\rangle|\Psi\rangle$, which in the
basis $|0\r$, $|1\r$ reads    
\begin{equation}
|\Psi\rangle |0\rangle \rightarrow |\Psi\rangle |0\rangle
+|0\r|\Psi\rangle =
2\alpha |00\rangle + \beta (|01\rangle +|10\rangle ).
\label{1.4}
\end{equation}
The vectors in the right-hand sides of Eqs.(\ref{1.3}) and (\ref{1.4}) 
are clearly not the same, no matter what
choice is made for $|v_{1}\rangle$ and $|v_{3}\rangle$.
Therefore, we need to search for devices which will 
produce approximate versions of the desired state or will
produce this state but with a probability which is less 
than one.

One way of creating a symmetrized state out of two independent
systems is by means of a measurement - that is the two
systems are optimally measured and their states are estimated. Based
on this estimation a two-particle entangled 
state is prepared.  If we begin with two qubits prepared so that 
one of the states is known ($|0\rangle$) while
the other is unknown ($|\Psi\rangle$), we need only estimate the 
state of one of the particles and this can be performed with a 
fidelity equal to 2/3 \cite{Massar,Derka}. The information gained from 
the optimal measurement is then used in the preparation procedure.
This is discussed in Section \ref{sec2a}.

We shall present quantum mechanical
entangling transformations which generate entangled states with
much higher fidelity than can be achieved by measuring the input
particles. 
In Section \ref{sec2b} we briefly discuss a {\em
probabilistic} symmetrization (entanglement) 
which can be realized via a controlled-SWAP
gate. The probability of success in this procedure 
is input-state dependent. In Section \ref{sec3} we present
the optimal input-state independent quantum entangler and we also
study the inseparabity of the outputs of this entangler.
In Section \ref{sec4} we show that the universal-NOT gate \cite{Buzek1}
can also serve as a very interesting entangling device.

\section{State-dependent symmetrization}
\label{sec2}
We shall first look at two examples of processes which produce
entangled states, for which the quality of the output depends on the
input state.  That is, these procedures work
better for some states than for others.  The first is perhaps the 
most obvious method, we simply measure the input state.  We shall
consider a more limited problem in this case, entangling an unknown 
with a known state.  The output state resulting from this procedure
is only an approximation to the desired one.  The second is a 
probabilistic method; the output when it is produced is ideal, but
the probability of successfully producing it is less than one.  In
this case we shall consider the full problem of entangling two
unknown states.

\subsection{Entanglement via measurement}
\label{sec2a}
Our task is to entangle  an input qubit in  an unknown state 
with a reference qubit in a known state $|0\rangle$. That is, 
we want to realize the symmetrization map 
$|0\r_{_{A}}|\Psi\r_{_{B}}\rightarrow |\Psi^{(id)}\r_{_{AB}}$ 
 with the output parameterized as
\be
|\Psi^{(id)}\r_{_{AB}}= 
\frac{2\cos\frac{\vartheta}{2}|00\r
+\sqrt{2} {\rm e}^{i\varphi}\sin\frac{\vartheta}{2} |+\r}
{\sqrt{2\left(1+\cos^2\frac{\vartheta}{2}\right)}},
\label{1.1.2}
\ee
The approach we will discuss here is as follows: firstly, the 
unknown single-qubit state $|\Psi\r$ is measured 
and then using the information gained
thereby  an approximate version of the desired output is constructed.
In order to specify this procedure in more detail, we must
describe what measurement is to be made and how its results
will be used to construct the output state.  The quality of
the output will be determined by calculating the fidelity
between the actual output and the desired output.  We shall first
examine a specific strategy and then find an upper bound on
the fidelity for a wide class of measurement-based procedures.
  
Our first measurement-based scenario can then be realized in 
the following way.  In the case of a single input qubit 
the {\em optimal} way to estimate
the state, is to measure it along a
randomly chosen direction in the two-dimensional
Hilbert space\cite{Massar,Derka}.  Therefore, the first step in 
implementing the measurement-based procedure is choosing a 
random vector $|\eta\rangle$, where
\be
|\eta\rangle = \cos\frac{\vartheta^{\prime}}{2} |0\rangle 
+{\rm e}^{i\varphi^{\prime}}\sin \frac{\vartheta^{\prime}}{2}|1\rangle ,
\label{1.1.5}
\ee
and measuring $|\Psi\rangle$ along it.  If the result
is positive, then
the output is taken to be $|\Phi\rangle_{_{AB}}$, and if
negative, the output is $|\tilde{\Phi}\rangle_{_{AB}}$,
where
\be
|\Phi\rangle_{_{AB}}
&=&\frac{|0\r|\eta\r +|\eta\r |0\r}
{\sqrt{2\left(1+\cos^2\frac{\vartheta^{\prime}}{2}\right)}} 
\nonumber
\\
&=& \frac{2\cos\frac{\vartheta^{\prime}}{2}|00\r
+\sqrt{2} {\rm e}^{i\varphi^{\prime}}\sin\frac{\vartheta^{\prime}}{2} |+\r}
{\sqrt{2\left(1+\cos^2\frac{\vartheta^{\prime}}{2}\right)}} 
\label{1.1.6}
\ee
and
\be
|\tilde{\Phi}\rangle_{_{AB}}
&=&\frac{|0\r|\eta^\perp\r +|\eta^\perp\r |0\r}
{\sqrt{2\left(1+\sin^2\frac{\vartheta^{\prime}}{2}\right)}} 
\nonumber
\\
&=& \frac{2{\rm e}^{-i\varphi^{\prime}}\sin\frac{\vartheta^{\prime}}{2}|00\r
-\sqrt{2} \cos\frac{\vartheta^{\prime}}{2} |+\r}
{\sqrt{2\left(1+\sin^2\frac{\vartheta^{\prime}}{2}\right)}} 
\label{1.1.7}
\ee
where the state $|\eta^\perp\r$ is the state orthogonal to 
$|\eta\r$, 
\be
|\eta^\perp\r= {\rm e}^{-i\varphi^{\prime}}
\sin\frac{\vartheta^{\prime}}{2} |0\rangle 
-\cos \frac{\vartheta^{\prime}}{2}|1\rangle .
\label{1.1.8}
\ee

For a particular orientation of the measurement apparatus, i.e. for 
the particular choice of the state $|\eta\r$ 
this measurement-based scenario gives the two-qubit 
output density matrix
\begin{equation}
{\rho}^{(out)}(\vartheta,\varphi | \vartheta^\prime,\varphi^\prime)
=|\langle\Psi |\eta\rangle |^{2}
|\Phi\rangle\langle\Phi |
+
|\langle\Psi 
|\eta^{\perp}\rangle |^{2}
|\tilde{\Phi}\rangle\langle \tilde{\Phi}|
 .\label{1.1.9}
\end{equation}
To get the final output density matrix one averages this over all
possible choices of the measurement (i.e. over all vectors
$|\eta\rangle$)
\begin{equation}
{\rho}^{(out)}(\vartheta,\varphi)
=\frac{1}{4\pi}\int_{0}^{2\pi}d\varphi^{\prime} 
\int_{0}^{\pi}d\vartheta^{\prime}\, \sin\vartheta^\prime \, 
{\rho}^{(out)}(\vartheta,\varphi| 
\vartheta^\prime,\varphi^\prime) .
\label{1.1.10}
\end{equation}
Finally, the fidelity can be found by computing the matrix
element of this density matrix in the ideal output state,
$|\Psi^{(id)}\rangle_{_{AB}}$,
\begin{equation}
\label{fidsp}
{\cal F}(\vartheta ,\varphi )=\langle\Psi^{(id)} |\rho^{(out)}
(\vartheta,\varphi)|\Psi^{(id)}\rangle .
\end{equation}
This fidelity depends on the input state, and this dependence
can be eliminated if we average over all input states
\begin{equation}
\label{fidgen}
\overline{{\cal F}}=\int d\Omega \, {\cal F}
(\vartheta ,\varphi ).
\end{equation}
This is the proper fidelity to use to judge how well our 
proposed strategy performs if we assume that all input states
are equally probable.  A more explicit expression for it is
\be
\overline{{\cal F}} &= &
\frac{1}{16\pi^2}\int_{0}^{2\pi}
\,d\varphi \,\int_{0}^{2\pi} d\varphi^\prime
\int_{0}^{\pi}\sin\vartheta d\vartheta\, 
 \int_{0}^{\pi} 
\sin\vartheta^\prime d\vartheta^\prime 
\nonumber \\
&\times &
\left[ |\l \eta|\Psi\r|^2 \ |\l \Psi|\Phi\r|^2 
  + |\l \eta^\perp|\Psi\r|^2 \ |\l \Psi
|\tilde{\Phi}\r|^2 \right],
\label{1.1.13}
\ee
Explicitly evaluating this integral we find
\be
\overline{{\cal F}} = 54 + 112\, (\ln 2)^2 
- 154.5 \ln 2\simeq  0.719 
\label{1.1.16}
\ee
which is a bit larger than $2/3$, the fidelity of 
the estimation of a state of a single qubit.

Let us now generalize this procedure.  We shall again begin by
choosing a random vector $|\eta\rangle$, but now according to a
distribution $q(\vartheta^{\prime},\varphi^{\prime})$, 
which we shall leave unspecified for now.
The output density matrix is taken to be either $\rho_{1}(\eta )$
if the measurement result is positive or $\rho_{0}(\eta )$ if
it is negative, where
\begin{equation}
\rho_{j}(\eta )=\int d\Omega^{\prime\prime}p_{j}
(\vartheta^{\prime\prime},
\varphi{\prime\prime}|\vartheta^{\prime},\varphi^{\prime})
|\Gamma (\vartheta^{\prime\prime },
\varphi^{\prime\prime})\rangle\langle\Gamma
(\vartheta^{\prime\prime },\varphi^{\prime\prime}) |,
\end{equation}
with $j=0,1$, and
\be
|\Gamma(\vartheta^{\prime\prime },\varphi^{\prime\prime})
\rangle_{_{AB}}
=\cos\frac{\vartheta^{\prime\prime}}{2}|00\rangle_{AB}
+ e^{i\phi^{\prime\prime}}\sin\frac{\vartheta^{\prime\prime}}
{2}|+\rangle_{AB} .
\ee
The conditional probabilities $p_{j}$ will
also be left unspecified; this allows us to consider a 
wide class of measurement-based strategies.  The output
density matrix, for a particular $|\eta\rangle$ is then
\begin{equation}
\rho (\eta )= |\langle\eta |\Psi\rangle |^{2}\rho_{1}(\eta )
+|\langle\eta^{\perp}|\Psi\rangle |^{2}\rho_{0}(\eta ) .
\end{equation} 
Averaging over $|\eta\rangle$ gives us the final output 
density matrix
\begin{equation}
\label{outgen}
\rho^{(out)}(\vartheta ,\varphi )
=\int d\Omega^{\prime}\rho (\eta )q(\vartheta^{\prime},
\varphi^{\prime}) ,
\end{equation}
and the fidelities for a specific input state and averaged over
all input states are given by Eqs. (\ref{fidsp}) and 
(\ref{fidgen}), respectively, but with $\rho^{(out)}$ computed
from Eq. (\ref{outgen}) instead of Eq. (\ref{1.1.10}).  In
particular we have that
\begin{eqnarray}
\overline{{\cal F}} = \int\int d\Omega^{\prime} d\Omega^{\prime
\prime}\sum_{j=0}^{1}f_{j}(\vartheta^{\prime\prime},\varphi^{
\prime\prime};\vartheta^{\prime},\varphi^{\prime}) P_{j}
(\vartheta^{\prime\prime},\varphi^{\prime\prime};
\vartheta^{\prime},\varphi^{\prime}) ,
\end{eqnarray}
where
\begin{equation}
P_{j}(\vartheta^{\prime\prime},\varphi^{\prime\prime};
\vartheta^{\prime},\varphi^{\prime}) = p_{j}
(\vartheta^{\prime\prime},\varphi^{\prime\prime}|
\vartheta^{\prime},\varphi^{\prime})q(\vartheta^{\prime},
\varphi^{\prime}) ,
\end{equation}
for $j=0,1$, and
\begin{eqnarray}
f_{0} &=& \int d\Omega \frac{1}{2(1+\cos^{2}
(\vartheta /2))}|2\cos\frac{\vartheta}{2}\cos\frac
{\vartheta^{\prime\prime}}{2} \nonumber \\
 & &+\sqrt{2}e^{i(\varphi^{\prime\prime}-\varphi )}\sin
\frac{\vartheta}{2}\sin\frac{\vartheta^{\prime\prime}}{2}|^{2}
|\langle\Psi |\eta^{\perp}\rangle |^{2} \nonumber \\
f_{1} &=& \int d\Omega \frac{1}{2(1+\cos^{2}
(\vartheta /2))}|2\cos\frac{\vartheta}{2}\cos\frac
{\vartheta^{\prime\prime}}{2} \nonumber \\
 & &+\sqrt{2}e^{i(\varphi^{\prime\prime}-\varphi )}\sin
\frac{\vartheta}{2}\sin\frac{\vartheta^{\prime\prime}}{2}|^{2}
|\langle\Psi |\eta\rangle |^{2} .
\end{eqnarray}
What we can now do is to find an upper bound for the fidelity,
$\overline{{\cal F}}$, for any distribution of the vector
$|\eta\rangle$ and any prescription for using the result of
the measurement along $|\eta\rangle$ to manufacture the
entangled state.  We note that for $j=0,1$
\begin{equation}
1=\int d\Omega^{\prime} \int d\Omega^{\prime\prime} P_{j}
(\vartheta^{\prime\prime},\varphi^{\prime\prime};
\vartheta^{\prime},\varphi^{\prime}) ,
\end{equation}
which implies that
\begin{equation}
\overline{{\cal F}}\leq \sup |f_{0}|+\sup |f_{1}| ,
\end{equation}
where the supremums are taken over the range $0\leq
\vartheta^{\prime},\vartheta^{\prime\prime} \leq \pi$ and 
$0\leq \varphi^{\prime}, \varphi^{\prime\prime} <2\pi $.

Our first task is to find explicit expressions for the
functions $f_{0}$ and $f_{1}$.  We have that
\end{multicols}
\vspace{-0.5cm}
\noindent\rule{0.5\textwidth}{0.4pt}\rule{0.4pt}{\baselineskip}
\widetext 
\begin{eqnarray}
f_{0} &=& d_{1}\cos^{2}\frac{\vartheta^{\prime\prime}}{2}
\sin^{2}\frac{\vartheta^{\prime}}{2}+d_{2}\cos^{2}\frac
{\vartheta^{\prime\prime}}{2}\cos^{2}\frac{\vartheta^{\prime}}
{2}+\frac{1}{2}d_{2}\sin^{2}\frac{\vartheta^{\prime\prime}}{2}
\sin^{2}\frac{\vartheta^{\prime}}{2} \nonumber \\
 & &+\frac{1}{2}d_{3}\sin^{2}\frac{\vartheta^{\prime\prime}}{2}
\cos^{2}\frac{\vartheta^{\prime}}{2}-\sqrt{2}d_{2}\cos
(\varphi^{\prime\prime}-\varphi^{\prime})\cos \frac
{\vartheta^{\prime\prime}}{2}\cos\frac{\vartheta^{\prime}}
{2}\sin\frac{\vartheta^{\prime\prime}}{2}\sin\frac
{\vartheta^{\prime\prime}}{2} \nonumber \\
f_{1} &=& d_{1}\cos^{2}\frac{\vartheta^{\prime\prime}}{2}
\cos^{2}\frac{\vartheta^{\prime}}{2}+d_{2}\cos^{2}\frac
{\vartheta^{\prime\prime}}{2}\sin^{2}\frac{\vartheta^{\prime}}
{2}+\frac{1}{2}d_{2}\sin^{2}\frac{\vartheta^{\prime\prime}}{2}
\cos^{2}\frac{\vartheta^{\prime}}{2} \nonumber \\
 & &+\frac{1}{2}d_{3}\sin^{2}\frac{\vartheta^{\prime\prime}}{2}
\sin^{2}\frac{\vartheta^{\prime}}{2} 
 +\sqrt{2}d_{2}\cos
(\varphi^{\prime\prime}-\varphi^{\prime})\cos \frac
{\vartheta^{\prime\prime}}{2}\cos\frac{\vartheta^{\prime}}
{2}\sin\frac{\vartheta^{\prime\prime}}{2}\sin\frac
{\vartheta^{\prime\prime}}{2}
\end{eqnarray}
\begin{multicols}{2} 
where
\begin{eqnarray}
d_{1} & = & 2\ln 2-1 \nonumber \\
 d_{2}& = & 3-4\ln 2 \\
d_{3} & = & 8\ln 2-5 .
\nonumber
\end{eqnarray}
From the above equations it is clear that in order to 
maximize $f_{0}$ we need to choose $\varphi^{\prime\prime}
-\varphi^{\prime}=\pi$ and to maximize $f_{1}$ we need to
choose $\varphi^{\prime\prime}-\varphi^{\prime}=0$.
Making these choices and simplifying the resulting
expressions we find that
\begin{eqnarray}
f_{0}(\vartheta^{\prime\prime},\pi;\vartheta^{\prime},0) &=&
\frac{1}{4}[1+c_{1}\cos\vartheta^{\prime\prime}-c_{2}\cos
\vartheta^{\prime\prime}\cos\vartheta^{\prime} \nonumber \\
 & &+c_{3}\sin\vartheta^{\prime\prime}\sin\vartheta^{\prime}]
\nonumber \\
f_{1}(\vartheta^{\prime\prime},0;\vartheta^{\prime},0) &=&
\frac{1}{4}[1+c_{1}\cos\vartheta^{\prime\prime}+c_{2}\cos
\vartheta^{\prime\prime}\cos\vartheta^{\prime} \nonumber \\
 & &+c_{3}\sin\vartheta^{\prime\prime}\sin\vartheta^{\prime}],
\end{eqnarray}
where
\begin{eqnarray}
c_{1} & = & 3-4\ln 2 \nonumber \\
 c_{2} & = & 12\ln 2-8  \\
c_{3} & = & \sqrt{2}(3-4\ln 2) .
\nonumber
\end{eqnarray}
These functions can now be maximized.  The maximum of $f_{0}$
occurs at $\vartheta^{\prime}=\pi$ and $\vartheta^{\prime\prime}
=0$, and the maximum of $f_{1}$ occurs when $\vartheta^{\prime}
=0$ and $\vartheta^{\prime\prime}=0$.  The maximum values of
both functions are the same and are approximately equal to
$0.386$.  This implies that the fidelity for this kind of
a measurement-based strategy must satisfy
\begin{equation}
\overline{{\cal F}}\leq 4\ln2 - 2 \cong 0.773 .
\end{equation}
As we shall see, a method which maintains quantum coherences
at all stages of the process can do better than this.

\subsection{Controlled-SWAP gate}
\label{sec2b}
We now begin with  systems $A$ and $B$ of the same physical origin. Their
pure states are  described by vectors 
in the $D$-dimensional Hilbert space $\mathcal{H}$, so that both together
are described by $\mathcal{H}\otimes \mathcal{H}$.  Let 
$\{|u_{j}\r|j=1,\ldots D\}$ be an orthonormal basis for $\mathcal{H}$.   
System $A$ is in the state
\begin{equation}
|\Psi\rangle_{_A}=\sum_{j=1}^{D}c_{j}|u_{j}\rangle_{_A} ,
\label{2.1}
\end{equation}
and system $B$ is in the state
\begin{equation}
|\Phi\rangle_{_B}=\sum_{j=1}^{D}d_{j}|u_{j}\rangle_{_B} .
\label{2.2}
\end{equation}
Our objective is to produce the (entangled) 
 symmetrized state [see Eq.~(\ref{1.1})]
\begin{equation}
\label{2.3}
|\Psi\rangle|\Phi\rangle+|\Phi\rangle|\Psi\rangle
\nonumber
\\
=\sum_{j=1}^{D}\sum_{k=1}^{D}(c_{j}d_{k}+c_{k}d_{j})|u_{j}\rangle
|u_{k}\rangle,
\end{equation}
(here we omit the normalization factor).

Recently Barenco et al. \cite{Barenco}
have shown that  the entanglement 
(symmetrization) of the form (\ref{1.1}) 
can be performed when the two input qubits interact
via  a controlled-SWAP
 (Fredkin) gate with an ancilla initially prepared in a specific state.
The entanglement is achieved when a conditional measurement is performed
on the ancilla. Exactly the same scenario can be used not only for qubits
but for arbitrary quantum systems. To show this we briefly review the
operation of the controlled-SWAP gate.

This gate has three inputs.  The first, the control bit, is a qubit.
The second and third are for $D$-dimensional systems.  The control bit 
is unaffected by the action of the gate.  If the control bit is 
$|0\rangle$, then the gate does nothing, i.e.\ the output state
is the same as the input state.  If the control bit is $|1\rangle$,
then the two $D$-dimensional states are swapped.  This can be
accomplished by the following explicit unitary transformation:
\begin{eqnarray}
|0\rangle |u_{j}\rangle |u_{k}\rangle & \rightarrow & 
|0\rangle |u_{j}\rangle |u_{k}\rangle;  \nonumber \\
|1\rangle |u_{j}\rangle |u_{k}\rangle & \rightarrow &
|1\rangle |u_{k}\rangle |u_{j}\rangle .
\label{2.4}
\end{eqnarray}
Summarizing, the action of our controlled-SWAP gate is,
\begin{eqnarray}
|0\rangle |\Psi\rangle|\Phi\rangle&\rightarrow &
|0\rangle |\Psi\rangle|\Phi\rangle;  \nonumber \\
|1\rangle |\Psi\rangle|\Phi\rangle &\rightarrow &
|1\rangle |\Phi\rangle|\Psi\rangle .
\label{2.5}
\end{eqnarray}

We now define the qubit states
\begin{eqnarray}
|v_+\rangle = \frac{1}{\sqrt{2}}(|0\rangle +|1\rangle );
\qquad
|v_-\rangle = \frac{1}{\sqrt{2}}(|0\rangle -|1\rangle ),
\label{2.6}
\end{eqnarray}
and take the input state of the controlled-SWAP gate to be
$|v_+\rangle |\Psi\rangle_{_A}|\Phi\rangle_{_B}$.  
Using the SWAP transformation (\ref{2.5}) we find that the 
output  state is
\begin{eqnarray}
|\Psi^{(out)}\rangle & = & \frac{1}{\sqrt{2}}\left(|0\rangle 
|\Psi\rangle
|\Phi\rangle+|1\rangle |\Phi\rangle |\Psi\rangle\right)
\nonumber \\
 & = &\frac{1}{2}|v_+\rangle \left(|\Psi\rangle|\Phi\rangle
+|\Phi\rangle|\Psi\rangle\right)
\\
&+&\frac{1}{2}|v_-\rangle
\left(|\Psi\rangle|\Phi\rangle-|\Phi\rangle
|\Psi\rangle\right) .
\nonumber
\end{eqnarray} 
If we now measure the qubit in the $|v_\pm\rangle$ basis we obtain 
the states $(|\Psi\rangle|\Phi\rangle\pm |\Phi\rangle
|\Psi\rangle)$ with probabilities $(1\pm |\langle \Psi |\Phi
\rangle|^{2})/2$, respectively.
As we see the probability of generation of a particular entangled state
explicitly depends on the (unknown) states of the two systems. In
particular, let us assume we begin with two orthogonal qubits, 
$|\Psi\r$ and $|\Psi^\perp\r$.  Then either of the maximally
entangled state, $(|\Psi\r |\Psi^\perp\r \pm
|\Psi^\perp\r |\Psi\r)/\sqrt{2}$ can prepared with probability
$1/2$. 

We  stress that 
the probability of the success in this entanglement (symmetrization)
procedure   is input-state dependent. 
In what follows our task will be to find a
``machine'' which entangles the input with a {\em constant} (i.e.
input-state independent) fidelity. This covariance property of the 
entangler with respect to unitary transformations performed
on the input qubits makes the entangler universal.

\section{Universal entanglers}
\label{sec3}
Suppose we again consider the problem of constructing
a device which will entangle a qubit in 
an arbitrary unknown  state $|\Psi\r=\alpha |0\r +\beta |1\r$
with a qubit in a  known, reference state, which we shall
take to be the basis state $|0\rangle$.  
Before we proceed further we have to specify properties
of the entangling map. In fact, we can consider two maps.
The symmetrization map
\be
{\cal S}:
|0\rangle_{_A}|\Psi\rangle_{_B} \rightarrow |\Psi^{(id)}\rangle_{_{AB}}=
N_{s}\left(|\Psi\rangle 
|0\rangle +|0\rangle |\Psi\rangle \right),
\label{3.1}
\ee
and the anti-symmetrization map
\be
{\cal A}:
|0\rangle_{_A}|\Psi\rangle_{_B} \rightarrow 
|\overline{\Psi}^{(id)}\rangle_{_{AB}}=
N_a\left(|\Psi\rangle 
|0\rangle -|0\rangle |\Psi\rangle \right), 
\label{3.2}
\ee
where $N_{a,s}$ are corresponding normalization factors.
As we have shown in the introduction perfect entanglers for arbitrary
unknown states cannot be constructed. So 
the task of the physically realizable 
symmetric (anti-symmetric)
entangler is to produce outputs as close as possible to the
ideally entangled states $|\Psi^{(id)}\rangle_{_{AB}}$
($|\overline{\Psi}^{(id)}\rangle_{_{AB}}$). 
In what follows we will quantify the quality of the performance of the
universal entangler with the help of the fidelity 
\begin{equation}
 {\cal F}:=\langle\Psi^{(id)}|\rho^{(out)}|\Psi^{(id)}\rangle.
\label{3.3}
\end{equation}
We shall impose the condition that the value of this fidelity 
does not depend
on the input.  The fidelity (\ref{3.3}) is a good measure
of the accuracy with which the entangler produces the desired
output state, but we would also like to evaluate the degree
of entanglement of the actual output state.  Here, however, 
we have a problem which is due to the fact that 
it is still not clear how to quantify the entanglement
of a quantum system which is in a mixed state. When a bipartite 
system
is in a pure state, then the von Neumann entropy of subsystems can 
serve as a measure of entanglement. In the case of impure states more 
sophisticated
measures are required (see for instance \cite{Vedral,Linden,Wootters}).

In terms of the basis vectors, the input
state is $\alpha |00\rangle +\beta |01\rangle$, and
the ideal output state in the case of symmetrization is 
\begin{equation}
|\Psi^{(id)}\rangle = \frac{
\left(2\alpha |00\rangle +\sqrt{2}\beta|+\rangle \right)
}{(4|\alpha |^{2} +2|\beta |^{2})^{1/2}} ,
\label{3.4}
\end{equation}
while in the case of the anti-symmetrization we have
\begin{equation}
|\overline{\Psi}^{(id)}\rangle = |-\r,
\label{3.5}
\end{equation}
where $|\pm\r$ are symmetric and anti-symmetric Bell states in the
given basis
\begin{equation}
|\pm \rangle = \frac{1}{\sqrt{2}}(|01\rangle \pm |10\rangle ).
\label{3.6}
\end{equation}
In what follows we will briefly discuss the anti-symmetric entangler 
and then we will concentrate on the symmetric entangler.

\subsection{Entanglement via anti-symmetrization}
\label{sec3a}
Recently Alber \cite{Alber}
studied a quantum entangler which takes as an input
a quantum-mechanical system prepared  
in an {\em unknown} pure state $|\Psi\rangle_{_{A}}$ and a reference
(known) state (let us say $|0\rangle_{_{A}}$ ) 
and at the output generates a two particle entangled state 
$\rho^{(out)}_{_{AB}}$ which is optimally entangled.
Alber imposed 
two constraints  on the output of the universal quantum entangler
\be
{\rm Tr}_{_A}\left[ \rho^{(out)}_{_{AB}}\right]=
{\rm Tr}_{_B}\left[ \rho^{(out)}_{_{AB}}\right]= \frac{\openone}{D}
\label{3.7}
\ee
and
\be
S\left[ \rho^{(out)}_{_{AB}}\right] \rightarrow {\rm minimum}.
\label{3.8}
\ee
Where $D$ is the dimensionality of the Hilbert space of the system $A$
($B$) and $S$ is the von Neumann entropy $S=-{\rm Tr}\rho\ln\rho$
associated with a given
density operator $\rho$.
The first condition corresponds to the requirement 
 that the subsystems at the output are in the maximally mixed state
while the second conditions guarantees that the whole system is as close
as possible to a pure two-particle state. Alber has found the solution 
for this problem. It turns out that 
the two-particle state which is produced by the optimal (with respect to the
above conditions), universal entangler is {\em independent} of
the input state $|\Psi\rangle$ and is equal to a maximally disordered
mixture of all possible anti-symmetric Bell states. In the case
of qubits ($D=2$) there is only one possible anti-symmetric  Bell
state $|-\r$. That is, Alber's machine realizes the anti-symmetric
entangler. We see that the universality of  Alber's entangler
means that all inputs are mapped to a single output (the anti-symmetric
Bell state $|-\r$), so the ideal output state is {\em a priori} known,
and one could instead build a device which just prepares the known 
output state.  In the antisymmetric entangler the information initially 
encoded in the qubit $A$ is completely  lost. But our task is different,
we want to redistribute the initial {\em unknown} information
encoded in the state of the qubit $A$, into the entangled state of two
qubits. Therefore we will analyze universal entanglement via
symmetrization, because the ideal state (\ref{3.4}) directly
contains information about the initial state of the qubit $A$. 
In other words, we consider the entangling procedure not only as
the way to generate the state with highest possible entanglement 
but also we require that this state contains as much information about
the input(s) as possible.

\subsection{Entanglement via symmetrization}
\label{sec3b}
Let us now construct a machine which entangles an unknown
state with the known state $|0\rangle$.
Taking into account the basic features of the symmetrization
transformation (\ref{3.1}) we can assume that the basis vectors 
transform as 
\begin{eqnarray}
\label{3.9}
|00\rangle|v_0\r &\rightarrow & |00\rangle |w_{0}\rangle +
|+\rangle |x_{0}\rangle \nonumber \\
|01\rangle|v_0\r  &\rightarrow & |00\rangle |w_{1}\rangle +
|+\rangle |x_{1}\rangle ,
\end{eqnarray}
where $|w_{0}\rangle$, $|w_{1}\rangle$, $|x_{0}\rangle$,
and $|x_{1}\rangle$ are states of the entangler itself.
The entangler is initially always prepared in the state $|v_0\r$.

We want to impose the condition that the fidelity between
the actual output state and the ideal output state be
{\em independent} of the state $|\Psi\rangle$, but before doing
so let us state the restrictions which unitarity places
on the machine vectors.  These are
\begin{eqnarray}
\label{3.10}
\|w_{0}\|^{2}+\|x_{0}\|^{2}&=&1 ; 
\nonumber
\\
\|w_{1}\|^{2}+\|x_{1}\|^{2} &=& 1;  
 \\
\langle w_{0}|w_{1}\rangle +\langle x_{0}|x_{1}\rangle &=& 0,
\nonumber
\end{eqnarray}
where $\|x\|^{2} \equiv \l x|x\r$.
We now calculate the output two-qubit density matrix
$\rho^{(out)}$ by using
the transformation in Eq. (\ref{3.9}) to find the full
output density matrix and then tracing out the machine degrees
of freedom. We then find the fidelity (\ref{3.3}) by taking the matrix
element of this density matrix in the ideal output state.
Our task is to find the machine vectors $|x_{j}\r$ and $|w_{j}\r$
($j=0,1$)
such that the fidelity $\cal{ F}$  does not depend on the input state 
$|\Psi\r$ and simultaneously is as close as possible to unity. 

We find
that if we choose $|x_{0}\rangle$ to be orthogonal to
each of the other machine vectors and $|w_{1}\rangle$ to
be orthogonal to $|x_{0}\rangle$ and $|w_{0}\rangle$, then
the output fidelity will be independent of the phases of
$\alpha$ and $\beta$.  Making these choices we find that
\begin{eqnarray}
\label{3.11}
{\cal F}& =& 
N^{-1} \left\{ 2|\alpha |^{4}
\|w_{0}\|^{2}+|\beta |^{4}\|x_{1}\|^{2} \right.
+|\alpha |^{2}|\beta |^{2}
\nonumber \\
 &\times &\left.
\left[\sqrt{2}(\langle w_{0}|x_{1}
\rangle +\langle x_{1}|w_{0}\rangle ) 
  +2\| w_{1}\|^{2}+\|x_{0}\|^{2}\right]\right\},
\end{eqnarray}
where $N=2|\alpha |^{2}+|\beta |^{2}$.

In order for this expression to be independent of 
$|\alpha|$ and $|\beta|$ it is necessary that the expression
in the curly brackets be proportional to
\begin{equation}
(2|\alpha |^{2}+|\beta |^{2})(|\alpha |^{2}+|\beta |^{2})
=2|\alpha |^{4}+3|\alpha |^{2}|\beta |^{2}+|\beta |^{4} .
\label{3.12}
\end{equation}
Comparing this expression to Eq. (\ref{3.11}) we see that
\begin{eqnarray}
\|w_{0}\| & = & \|x_{1}\| \nonumber \\
3 \| w_{0}\|^{2}
& = & \sqrt{2} (\langle x_{1}|w_{0}\rangle +\langle w_{0}| 
x_{1}\rangle )+ 2\| w_{1}\|^{2}+\| x_{0}\|^{2}.
\label{3.13}
\end{eqnarray}
If these conditions are satisfied, then the fidelity is
simply equal to $\| w_{0}\|^{2}$, so that we want to 
make this quantity as large as possible.  If we now
make use of the unitarity conditions and the two 
equations above, we find that 
\begin{equation}
\label{w00}
1-\frac{2}{3}\sqrt{2}\cos\mu = \frac{1-\|w_{0}\|^{2}}
{\|w_{0}\|^{2}} ,
\label{3.14}
\end{equation}
where
\begin{equation}
\cos\mu = \frac{\langle x_{1}|w_{0}\rangle +\langle w_{0}|
x_{1}\rangle}
{2\| w_{0}\|^{2}} .
\label{3.15}
\end{equation}
From Eq. (\ref{w00}) we see that 
$\| w_{0}\|^{2}$ will be a maximum when $\cos\mu = 1$, 
which implies that $|w_{0}\rangle$ and $|x_{1}\rangle$
are parallel.  When this condition is satisfied, we find
that
\begin{equation}
{\cal F}=\| w_{0}\|^{2}=\frac{9+3\sqrt{2}}{14} ,
\label{3.16}
\end{equation}
which gives $0.946$ as the approximate value of the
fidelity. This means that the output state $\rho^{(out)}$ is
indeed very close to the ideal state, and it should be remembered
that this fidelity is the same for all input states.

We can summarize our results for the machine vectors
as follows.  From the above analysis we see that we
can take the machine state space to be {\em three}
dimensional.  Define
\begin{equation}
\cos\theta = \left[ \frac{9+3\sqrt{2}}{14}\right]^{1/2};
\qquad
\sin\theta = \left[ \frac{5-3\sqrt{2}}{14}\right]^{1/2},
\label{3.17}
\end{equation}
and let $\{|v_{j}\rangle |j=1,\ldots 3\}$ be an 
orthonormal basis for the machine vector space.  
We then have
\begin{eqnarray}
\label{3.18}
|w_{0}\rangle &=& \cos\theta |v_{1}\rangle \nonumber \\
|w_{1}\rangle &=& \sin\theta |v_{2}\rangle \nonumber \\
|x_{0}\rangle &=& \sin\theta |v_{3}\rangle \\
|x_{1}\rangle &=& \cos\theta |v_{1}\rangle ,\nonumber
\end{eqnarray}  
and our transformation in terms of basis vectors 
becomes
\begin{eqnarray}
|00\rangle |v_0\r &\rightarrow & \cos\theta 
|00\rangle|v_{1}\rangle 
+\sin\theta  |+\rangle |v_{3}\rangle ;
\nonumber
\\
|01\rangle |v_0\r &\rightarrow & \sin\theta 
|00\rangle|v_{2}\rangle 
+\cos\theta  |+\rangle |v_{1}\rangle.
\label{3.19}
\end{eqnarray}
By  construction this is the {\em optimal} 
entangling transformation which entangles an unknown
pure state with a known reference state.

Alternatively, for $|\Psi\r = \alpha|0\r +\beta |1\r$
we can rewrite this transformation
in the form
\begin{eqnarray}
|0\rangle |\Psi\rangle |v_0\r
&\rightarrow & \cos\theta (\alpha |00\r +\beta |+\r) |v_{1}\rangle 
\nonumber \\
 & & +\sin\theta \left(\alpha |+\rangle |v_{3}\rangle +
\beta|00\rangle |v_{2}\rangle \right) .
\label{3.20}
\end{eqnarray}
When  the trace over the entangler is performed we obtain the 
density operator 
 $\rho_{_{AB}}^{(out)}$ describing the two qubits  $A$ and
$B$ at the output of the quantum entangler
\begin{eqnarray}
\label{3.21}
\rho^{(out)}_{AB}
&=&( |\alpha|^2 \cos^2\theta +|\beta|^2 \sin^2\theta) |00\r\l 00|
\nonumber
\\
&+&( |\alpha|^2 \sin^2\theta +|\beta|^2 \cos^2\theta) |+\r\l +|
\\
&+& \cos^2\theta ( \alpha \beta^* |00\r\l +|
+\alpha^* \beta |+\r\l 00|)
\nonumber
\end{eqnarray}
It is important to stress that the fidelity (\ref{3.3}) 
associated with the output state (\ref{3.20}) is input state
independent.

\subsection{Remarks}
\label{sec3c}
Throughout this paper we have utilized the fidelity (\ref{3.3}) as the
measure of the performance of the quantum entangler. The universality
(covariance) of the entangler is expressed in the fact that the
value of the fidelity ${\cal F}$ is equal for all input states.
We note that this covariance constraint is equivalent to the
requirement that the Bures distance  
\cite{Bures} defined as 
\begin{eqnarray}
d_B({\rho}_1,{\rho}_2)
=\sqrt{2}\left(1 -{\rm Tr}\sqrt{\hat{\rho}_1^{1/2}\hat{\rho}_2
\hat{\rho}_1^{1/2}}\right)^{1/2},
\label{3.22}
\end{eqnarray}
between the ideal state $|\Psi^{(id)}\r$ and the output of the
entangler $\rho^{(out)}_{_{AB}}$ is constant. In our particular
case we find the Bures distance to be
\be
d_B= 2 \sin(\theta/2) \simeq 0.0541,
\label{3.23}
\ee
for all inputs. This distance is very small indeed.
It is important to note that the Hilbert-Schmidt norm
\be
d_{HS}(\rho_1,\rho_2)=\left[ {\rm Tr}(\rho_1 - \rho_2)^2\right]^{1/2},
\label{3.24}
\ee
which in our case can be expressed as
\be
d_{HS}=\left[ 1-2{\cal F} + {\rm Tr}\left(\rho^{(out)}_{_{AB}}\right)^2 
\right]^{1/2},
\label{3.25}
\ee
is {\em not} input-state independent because 
${\rm Tr}\left(\rho^{(out)}_{_{AB}}\right)^2 $ depends on the initial state.
This  is closely related to the fact that the von Neumann entropy
of the state $\rho^{(out)}_{_{AB}}$ is state dependent (see below).

\subsection{Inseparability of the output qubits}
\label{sec3d}
\begin{figure}
\centerline {\epsfig{width=8.0cm,file=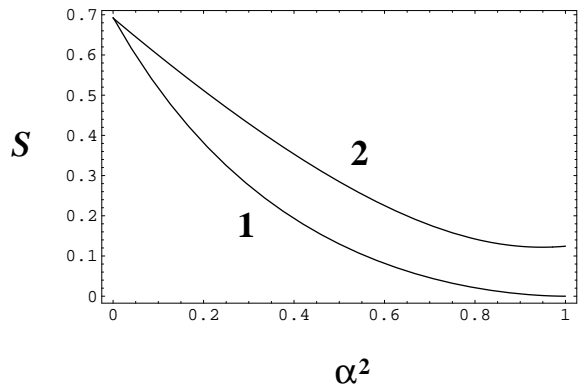}}
\begin{narrowtext}
\bigskip      
\caption{ 
The von Neumann entropy of the single-qubit state $\rho_{_A}$ when
the two-qubit system is in an ideally entangled state $|\Psi^{(id)}\r$
(line 1) and when 
 the output state $\rho_{_{AB}}^{(out)}$ is  given by Eq.(\ref{3.21})
(line 2).
In both cases we assume $\alpha$ and $\beta$ to be real.
}
\label{ent_fig1}
\end{narrowtext}
\end{figure}

We note that the entanglement between the two qubits
prepared in the state $|\Psi^{(id)}\r$ depends on the particular
form of the state $|\Psi\r=\alpha|0\r+\beta|1\r$. Because $|\Psi^{(id)}\r$
is a pure state we can quantify the degree of entanglement via the
von Neumann entropy $S$ of one of the two qubits under consideration,
i.e. $S_A=-{\rm Tr}[\rho_A\ln\rho_A]$ (obviously $S_A=S_B$).
For $\alpha=1$ the entropy is equal to zero, which corresponds 
to a completely
disentangled state (we note that in this case $|\Psi^{(id)}\r =|0\r|0\r$).
The entropy takes the maximal value $S=\ln 2$ for $\alpha=0$ when
$|\Psi^{(id)}\r =(|0\r|1\r+|1\r|0\r)/\sqrt{2}$.
We plot this entropy in Fig.~\ref{ent_fig1} (see line 1). 
The entropy of the individual particle (qubit) at the output of
the entangler, i.e. $\rho_{_{A}}^{(out)}={\rm Tr}\rho_{_{AB}}^{(out)}$
is always larger than in the ideal case (see line 2 in
Fig.~\ref{ent_fig1}). Nevertheless, for the case $\alpha=0$ we have
in this case $S(\alpha=0)=0.998 \ln 2$, i.e. this entropy is very close
to the entropy of a qubit in the ideal case. Unfortunately, this
entropy in the case of an impure two-particle state cannot be used
as a measure of entanglement. 
%%%%%%%%%%%%%%%%%%%%%%%%%%%%%%%%%%%%%%%%%%%%%%%%%%%%%%%%%%%%%%%%%%%%%%%%%%
%%%%%%%%%%%%%%%%%%%%%%%%%%%%%%%%%%%%%%%%%%%%%%%%%%%%%%%%%%%%%%%%%%%%%%

\begin{figure}
\centerline {\epsfig{width=8.0cm,file=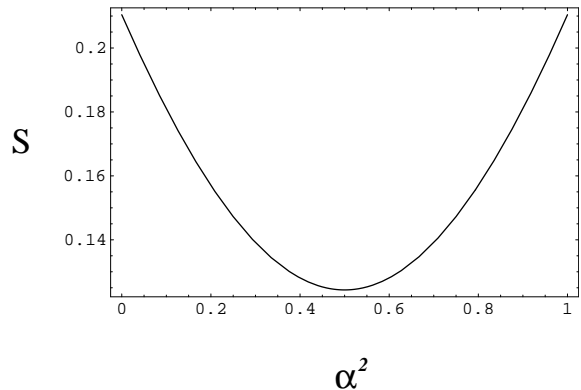}}
\begin{narrowtext}
\bigskip      
\caption{ 
The von Neumann entropy of the two-qubit state $\rho_{_{AB}}^{(out)}$ 
at the output of the entangler [see Eq.~(\ref{3.21})]
as a function of $\alpha^2$. 
We assume $\alpha$ and $\beta$ to be real.
}
\label{ent_fig2}
\end{narrowtext}
\end{figure}

It is interesting to find the entropy of the two-particle state 
$\rho_{_{AB}}^{(out)}$ at the output of the entangler as a function
of the initial state (in the ideal case the two-particle system is
always considered to be in a pure state with $S=0$). We plot this
entropy in Fig.~\ref{ent_fig2}.
We see that the total entropy of the output is state-dependent and it takes
the minimal value for $\alpha^2=1/2$. Therefore the entropy of the
subsystems does not indicate whether they are entangled.
%%%%%%%%%%%%%%%%%%%%%%%%%%%%%%%%%%%%%%%%%%%%%%%%%%%%%%%%%%%%%%%%%%%%%%
%%%%%%%%%%%%%%%%%%%%%%%%%%%%%%%%%%%%%%%%%%%%%%%%%%%%%%%%%%%%%%%%%%%%%

We need to check whether the two qubits $A$ and $B$
at the output  are indeed quantum-mechanically
entangled. 
Quantum-mechanical entanglement of two qubits 
formally means that the density operator of these two qubits is
represented by an inseparable matrix (see \cite{Peres1}).
It follows from the Peres-Horodecki theorem that
\cite{Peres2,Horodecki1} the necessary and sufficient condition
of inseparability of  the two-qubit density matrix $\rho_{_{AB}}$ 
is that the corresponding 
partially transposed  matrix $\rho^{T_2}_{_{AB}}$ has at least one 
 negative eigenvalue.

\begin{figure}
\centerline {\epsfig{width=8.0cm,file=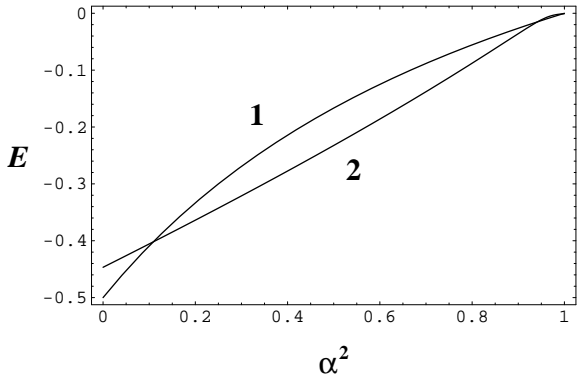}}
\begin{narrowtext}
\bigskip      
\caption{ 
Here we plot the negative eigenvalue Eq.~(\ref{3.26})
of the partially transposed
matrix of the density operator $\rho^{(ideal)}_{_{AB}}$ when the state
$|\Psi\r$ has real amplitudes $\alpha$ and $\beta$ (see line 1).
The negative eigenvalue of  
the partially transposed matrix associated with 
 the density operator $\rho^{(out)}_{_{AB}}$ given by Eq.~(\ref{3.21})
as functions of $\alpha$ is presented by line 2. (We assume $\alpha$
and $\beta$ to be real.) 
}
\label{ent_fig3}
\end{narrowtext}
\end{figure}

For instance, let us consider the state 
$|\Psi\r=\alpha|0\r+\beta|1\r$
with real amplitudes $\alpha$ and $\beta$. 
The partially transposed matrix corresponding to the
state $|\Psi^{(id)}\r$ given by Eq.(\ref{3.4})
has one negative eigenvalue 
\begin{eqnarray}
E(\alpha)=\frac{\alpha^2-1}{2(\alpha^2+1)}.
\label{3.26}
\end{eqnarray}
We plot this eigenvalue in Fig.~\ref{ent_fig3} (see line 1). We see that the
eigenvalue
is negative for all values of $\alpha$ except $\alpha=1$ when 
$|\Psi^{(id)}\r =|0\r|0\r$. The minimal value of the eigenvalue
is achieved for $\alpha=0$ when the two qubits are in the maximally entangled
state $(|01\rangle+|10\rangle)/\sqrt{2}$.
%%%%%%%%%%%%%%%%%%%%%%%%%%%%%%%%%%%%%%%%%%%%%%%%%%%%%%%%%%%%%%%%%%%%%
%%%%%%%%%%%%%%%%%%%%%%%%%%%%%%%%%%%%%%%%%%%%%%%%%%%%%%%%%%%%%%%%%%%%%%%%%

Now we utilize the Peres-Horodecki theorem to check whether the
state $\rho^{(out)}_{_{AB}}$ given by Eq.~(\ref{3.21}) describes
an entangled state of two qubits. Firstly, we find that 
the partially
transposed matrix corresponding to the density operator (\ref{3.21})
has one eigenvalue which is negative for all values of $\alpha$
(here we assume $\alpha$ and $\beta$ to be real). In particular,
this eigenvalue for $\alpha=0$ is
\begin{eqnarray}
E(\alpha=0)=\frac{1}{2}\left[\cos^2\theta-\left(\cos^4\theta
+\sin^4\theta\right)^{1/2}\right],
\label{3.27}
\end{eqnarray}
which is the minimal value ($\simeq -0.447$)
of the negative eigenvalue. On the other
hand the maximal value ($\simeq -0.001$)
is attained for $\alpha=1$
\begin{eqnarray}
E(\alpha=1)=\frac{1}{2}\left[\sin^2\theta-\left(\cos^4\theta
+\sin^4\theta\right)^{1/2}\right].
\label{3.28}
\end{eqnarray}
The complete dependence of $E(\alpha)$ is shown in Fig.~\ref{ent_fig3}.
From this figure we clearly see that the output density operator
is inseparable for an arbitrary input considered in this Section.
We note, that if the entanglement is measured in
terms of the tangle as introduced by Wootters \cite{Wootters} 
then the negative eigenvalues $E$ of the partially transposed density
operators  perfectly reflect the degree of 
entanglement between the two qubits in our cases.

By construction the fidelity of the entangler 
in this case is constant
but the actual degree of entanglement is state-dependent. 
This suggests that it would be interesting
to find an entangler, whose output states have the same
degree of entanglement irrespective of the input, yet still 
carry information about the input.

\section{Entanglement via universal NOT gate}
\label{sec4}
Even though the negative eigenvalue of the partially transposed density
matrix cannot be directly used as the measure of entanglement, we
see that the degree of entanglement between two qubits generated
in the entangler (\ref{3.19}) depends on the input state. In what follows
we describe a different type of the entangler, which out of a single
qubit $|\Psi\r$ generates a two-qubit state as close
as possible to the state 
\begin{eqnarray}
|\Psi\r\rightarrow |\{\Psi,\Psi^\perp\}\r\equiv
(|\Psi\r|\Psi^\perp\r+|\Psi^\perp\r|\Psi\r)/\sqrt{2}.
\label{4.1}
\end{eqnarray}
We will present an
 entangler which  not only
produces the state which is as close as possible
to the ideal state $|\{\Psi,\Psi^\perp\}\r$ 
but also has the property
that the fidelity does not
depend on the input state. In addition, the degree of entanglement also 
does not depend on the input. This type of the entangler implicitly assumes 
creation of the state $|\Psi^\perp\rangle$ from the input $|\Psi\rangle$.
That is, we face the problem of creating an orthogonal state from 
unknown input. 

It is not a problem to
complement a classical bit, i.e. to change the value of a bit, a
$0$ to a $1$ and vice versa.  This is accomplished by a NOT gate.
Complementing a qubit, however, is another matter.  The complement
of a qubit $|\Psi\rangle$ is the qubit $|\Psi^{\perp}\rangle$
which is orthogonal to it.  But  it is not possible to build a 
device which
will take an {\it arbitrary} (unknown)
 qubit and transform it into the qubit
orthogonal to it.
As shown in Ref.~ \cite{Buzek1}
the ideal universal-NOT (U-NOT) operation corresponds to  the
{\it inversion of the Bloch (Poincar\'e) sphere}.
This inversion preserves angles (related in a simple way
to the scalar product $\vert\l\Phi |\Psi\r\vert$ of rays), so by
Wigner's Theorem the ideal U-NOT must be implemented either by a
unitary or by an anti-unitary operation. Unitary operations
correspond to proper rotations of the Poincar\'e sphere, whereas
anti-unitary operations correspond to orthogonal transformations
with determinant $-1$. Clearly, the U-NOT operation is of the
latter kind, and an anti-unitary operator $\Theta$ (unique up to a
phase) implementing it is
\begin{equation}
\Theta\bigl(\alpha\vert0\r+ \beta\vert1\r\bigr)
   =\beta^{\ast}\vert0\r-\alpha^{\ast}\vert1\r.
\label{theta}
\end{equation}
The difficulty with anti-unitarily implemented symmetries is that
they are not completely positive, i.e., they cannot be applied to
a small system, leaving the rest of the world alone.

Because we cannot design a perfect Universal-NOT gate,  we
have introduced in Ref.\cite{Buzek1} an approximate {\em optimal}
U-NOT gate (an analogous spin-flip operation has recently been 
introduced by Gisin and Popescu \cite{GisinPopescu}).
This device takes as an input the qubit $A$ in the
state $|\Psi\rangle$ and generates at the output a qubit in
a mixed state as close as possible to the orthogonal state
$|\Psi^\perp\rangle$. The role of the U-NOT gate is played
by two additional (ancilla) qubits $B$ and $C$. So, all together
the transformation involves three qubits and it can be explicitly written
as 
\begin{eqnarray}
|\Psi\r_{_A}|X\r_{_{BC}}& \rightarrow&
\gamma_0|\Psi,\Psi\r_{_{AB}}|\Psi^\perp\r_{_C}
\nonumber
\\
& & +
\gamma_1
|\{\Psi,\Psi^\perp\}\r_{_{AB}}
|\Psi\r_{_C},
\label{4.2}
\end{eqnarray}
where $|X\r_{_{BC}}$ is the initial state of the U-NOT gate;
 $\gamma_0=\sqrt{2/3}$ and $\gamma_1=-\sqrt{1/3}$.
In this particular transformation the  qubit $C$ at the output is
in the state which is as orthogonal as possible to the  input state.
The fidelity of this transformation is input-state independent and
is equal to ${\cal F}=2/3$.  

\subsection{U-NOT as the entangler}
It is interesting to note that 
the two-qubit state $\rho^{(out)}_{_{AB}}$ at the output of the U-NOT gate
(\ref{4.2}) has the form
\begin{eqnarray}
\rho^{(out)}_{_{AB}}=
\gamma_1^2|\{\Psi,\Psi^\perp\}\r\l\{\Psi,\Psi^\perp\}|
+\gamma_0^2|\Psi\Psi\r\l\Psi\Psi|.
\label{4.3}
\end{eqnarray}
The mean fidelity between the state $\rho^{(out)}_{_{AB}}$ and the
ideal output (\ref{4.1}) is input-state independent and takes the value
${\cal F}=1/3$. This again corresponds to the fact that the Bures
distance between the actual output of the entangler and the ideal output
is input state independent and equal to $d_B=(2-2/\sqrt{3})^{1/2}$. 
We can easily check that the partially transposed matrix corresponding to
the density operator (\ref{4.3}) has one negative eigenvalue
$E=(2-\sqrt{5})/6$ which is constant and does not depend on the initial
input state $|\Psi\r$.

We note that the Universal NOT gate (\ref{4.2}) acts also a quantum cloner,
i.e. the two qubits $A$ and $B$ are the optimal clones of the input
(for details see Refs.\cite{Buzek2,Gisin1}).
It is the optimality of the transformation (\ref{4.2}) with respect to
cloning and the generation of the optimally orthogonal state (i.e. the
universal NOT gate) which indicates that the transformation (\ref{4.2}) 
also serves as the optimal universal entangler.

\subsection{Proof of optimality}
Our proof of the optimality of the entangler (\ref{4.1}) via the U-NOT
gate is based on the recent idea of  Gisin \cite{Gisin2,Gisin3} that the
impossibility of instantaneous signaling generates upper bounds on 
the fidelity of particular quantum-mechanical processes. 
To be more specific, we  have shown earlier that the impossibility
of the ideal (perfect) entangler is due to the linearity of quantum
mechanics. On the other hand, another consequence of the linearity of
quantum mechanics is the fact that the entangled quantum-mechanical
states cannot be used for super-luminal communication. Gisin \cite{Gisin2}
has shown that this no-signaling constraint implies bounds on the
fidelity of universal cloning and the universal U-NOT gate. In the 
case of cloning the bound on fidelity is ${\cal F}=5/6$, while
in the case of the U-NOT gate the bound is ${\cal F}=2/3$. We note that the 
transformation (\ref{4.2}) achieves both these bounds when used as the
cloner or the U-NOT gate, respectively.
Recently Alber \cite{Alber} used this idea of Gisin to prove that the
upper bound in the fidelity of the anti-symmetric entangling  
is equal to unity. The no-signaling constraint can also be used to 
derive an upper bound on the fidelity of the entangling operation
given in Eq. (\ref{4.1}) \cite{Gisin3}. 
We will present a proof, which is based on the methods developed 
in reference \cite{Gisin2}, 
that this upper bound is ${\cal F}=1/3$, which means
that the U-NOT gate (\ref{4.2}) serves as the {\em optimal} universal 
entangler in the sense of Eq. (\ref{4.1}).

We consider a process in which a single particle input
state is mapped into a two particle output state.  The
input state can be represented as
\begin{equation}
\rho^{(in)}(\vec{m})=\frac{1}{2}(\openone+\vec{m}\cdot\vec{\sigma} ) ,
\end{equation}
where $\vec{m}$ is a real vector whose length is less than
or equal to unity.
The most general two-particle output state, which is
hermitian and has a trace equal to one, can be expressed as
\be
\rho^{(out)}(\vec{m})
&=& \frac{1}{4}[\openone+\vec{a}\cdot\vec{\sigma}
\otimes \openone+\openone\otimes\vec{b}\cdot\vec{\sigma}
\nonumber
\\
&+&\sum_{j,k=x,y,z}
t_{jk}\sigma_{j}\otimes\sigma_{k} ] ,
\ee
where $\vec{a}$, $\vec{b}$, and $t_{jk}$ are functions of
$\vec{m}$.  The requirement that the reduced density matrixes
of the two output particles be the same, which we shall
impose, implies that $\vec{a}=\vec{b}$.

We now want to impose the requirement of covariance.  This
means that if $\rho^{(in)}(\vec{m})$ is mapped onto 
$\rho^{(out)}(\vec{m})$, and if $u$ is a matrix in $SU(2)$, then 
the input state $u\rho^{(in)}(\vec{m})u^{-1}$ will be mapped
onto the output state $u\otimes u\rho^{(out)}(\vec{m})u^{-1}
\otimes u^{-1}$.  Another way of stating this condition is
obtained by noting
that if we express $u$ as
\begin{equation}
u=\exp(-i\theta\hat{e}\cdot\vec{\sigma}/2) ,
\end{equation}
where $\hat{e}$ is a unit vector corresponding to the rotation
axis and $\theta$ is the rotation angle, then
\begin{equation}
u(\vec{m}\cdot\vec{\sigma})u^{-1}=\vec{m}^{\prime}\cdot
\vec{\sigma},
\end{equation}
where $\vec{m}^{\prime}=R(\hat{e},\theta )\vec{m}$.  The
rotation matrix, $R(\hat{e},\theta )$, is the 
$3\times 3$ matrix which rotates a vector about
the axis $\hat{e}$ by an angle $\theta$, and it is given
explicitly by
\begin{equation}
R(\hat{e},\theta )=\exp (\theta \hat{e}\cdot\vec{K}) ,
\end{equation}
where
\begin{eqnarray}
K_{x}=\left(\begin{array}{ccc} 0 & 0 & 0 \\ 0 & 0 & -1 \\
0 & 1 & 0 \end{array}\right) ,\nonumber \\
K_{y}=\left(\begin{array}{ccc} 0 & 0 & 1 \\ 0 & 0 & 0 \\
-1 & 0 & 0 \end{array}\right) ,\nonumber \\
K_{z}=\left(\begin{array}{ccc} 0 & -1 & 0 \\ 1 & 0 & 0 \\
0 & 0 & 0 \end{array}\right) .
\end{eqnarray}
We have that 
\begin{equation}
u\rho^{(in)}(\vec{m})u^{-1}=\rho^{(in)}(R\vec{m}) ,
\end{equation}
which will be mapped to $\rho^{(out)}(R\vec{m})$, so
that the covariance condition can now be expressed as
\begin{equation}
\label{cov}
\rho^{(out)}(R\vec{m})=u\otimes u\rho^{(out)}(\vec{m})u^{-1}
\otimes u^{-1} .
\end{equation}

Now let us examine the consequences of this relation.  We
shall first consider the terms linear in $\vec{\sigma}$ and
let $R$ be a rotation about $\vec{m}$ by a very small angle
$\theta$.  We have that 
\begin{equation}
\label{Ra}
\vec{a}(R\vec{m}) =R\vec{a}(\vec{m}) ,
\end{equation}
which for our choice of rotation becomes
\begin{equation}
\vec{a}(\vec{m})=(\openone+\theta \hat{m}\cdot\vec{K})\vec{a}(\vec{m}) ,
\end{equation}
or
\begin{equation}
\hat{m}\cdot\vec{K}\vec{a}(\vec{m}) =0 ,
\end{equation}
where $\hat{m}$ is a unit vector in the direction of $\vec{m}$.
This implies that $\hat{m}\times\vec{a}(\vec{m})=0$, so that
$\vec{a}(\vec{m})$ is parallel to $\vec{m}$, and we can write
$\vec{a}(\vec{m})=a(\vec{m})\vec{m}$.  If we now substitute
this result back into Eq. (\ref{Ra}) and consider a general
rotation $R$, we have that
\begin{equation}
a(R\vec{m})=a(\vec{m}) .
\end{equation}
This implies that $a(\vec{m})$ is a constant, which, following
\cite{Gisin2}, we shall denote by $\eta$.

Now let us see what covariance implies about the terms
quadratic in $\vec{\sigma}$.  Application of the covariance
condition, Eq. (\ref{cov}), to these terms gives
\begin{equation}
t_{jk}(R\vec{m})=\sum_{j^{\prime},k^{\prime}}R_{jj^{\prime}}
R_{kk^{\prime}}t_{j^{\prime}k^{\prime}}(\vec{m}) .
\end{equation}
If we again choose $R$ to be a rotation about $\vec{m}$ 
by a small angle $\theta$, we find the condition
\begin{equation}
0=\sum_{j^{\prime}}(\hat{m}\cdot \vec{K})_{jj^{\prime}}t_{j^{\prime}k}
(\vec{m})+\sum_{k^{\prime}}(\hat{m}\cdot \vec{K})_{kk^{\prime}}
t_{jk^{\prime}}(\vec{m}) .
\end{equation}
If we choose $\vec{m}$ to be in the $z$ direction, in
particular $\vec{m}=\hat{z}$, we find, as did Gisin,
that $t_{xx}=t_{yy}$, $t_{xy}=-t_{yx}$, and $t_{xz}=
t_{zx}=t_{yz}=t_{zy}=0$, where all of these are
evaluated at $\vec{m}=\hat{z}$.  We now want to impose
the no signaling condition
\begin{equation}
\label{nosig}
\rho^{(out)}(\hat{z})+\rho^{(out)}(-\hat{z})=\rho^{(out)}(\hat{x})
+\rho^{(out)}(-\hat{x}) ,
\end{equation}
and to do so we need to find all of the density matrixes
in the above equation in terms of $t_{jk}(\hat{z})$.
This can be done by applying the covariance condition,
Eq. (\ref{cov}), to $\rho^{(out)}(\hat{z})$ and making the
proper choice of $R$.  When these results are substituted
into Eq. (\ref{nosig}) we find that $t_{xx}(\hat{z})=
t_{yy}(\hat{z})=t_{zz}(\hat{z})$, and we shall designate
this common value by $t(\hat{z})$.  We then have that
\begin{equation}
\rho^{(out)}(\hat{z})=\frac{1}{4}\left( \begin{array}{cccc}
1+2\eta +t & 0 & 0 & 0 \\ 0 & 1-t & 2(t+it_{xy}) & 0 \\
0 & 2(t-it_{xy}) & 1-t & 0 \\ 0 & 0 & 0 & 1-2\eta +t
\end{array}\right) .
\end{equation}
The basis in which the matrix is expressed is $\{ |+\hat{z},
+\hat{z}\rangle , |+\hat{z},-\hat{z}\rangle ,|-\hat{z},
+\hat{z}\rangle ,|-\hat{z},-\hat{z}\rangle \}$, where
$\sigma_{z}|\pm\hat{z}\rangle =\pm |\hat{z}\rangle$.
This matrix must be positive, which implies that the 
eigenvalues
\begin{equation}
\frac{1}{4}(1\pm 2\eta +t); \qquad \frac{1}{4}(1-t\pm
2\sqrt{t^{2}+t^{2}_{xy}}) 
\end{equation}
must be nonnegative.

For an input state $\rho^{(in)}(\hat{z})$ our desired output
state is $(|+\hat{z},-\hat{z}\rangle +|-\hat{z},
+\hat{z}\rangle )/\sqrt{2}$, and this implies that the
fidelity of $\rho^{(out)}$ is
\begin{equation}
{\cal F}=\frac{1+t}{4} .
\end{equation}
This is clearly maximized when $t$ is as large as possible,
and examining the eigenvalues of $\rho^{(out)}$, this 
happens when $t_{xy}=0$ and $t=1/3$.  Substituting
this into the expression for the fidelity, we see that the
maximum fidelity is $1/3$.  This means that the
no-signaling constraint specifies  the upper bound on 
the fidelity of the symmetric entangling which is exactly 
the same one as achieved by the U-NOT gate. This proves that 
the entangling via the U-NOT gate is optimal.

\subsection{Remark}
We note that using the universal NOT gate one can also produce 
an entangled state of the form (\ref{3.1}). Specifically,
the U-NOT gate  
 allows Charlie (C) to produce an entangled state,
consisting of $|\Psi\r$ and one of two known states, which is
shared by Alice (A) and Bob (B).  In order to see how
this can be accomplished it is useful to express the state
on the right-hand side of Eq. (\ref{4.2}) as
\begin{eqnarray}
\sqrt{\frac{1}{3}}(|\Psi\r_{_A}|\Phi_{-}\r_{_{BC}} +|\Psi\r_{_B}
|\Phi_{-}\r_{_{AC}}),
\end{eqnarray}
where
\begin{eqnarray}
|\Phi_{-}\r = \frac{(|\Psi\r |\Psi^{\perp}\r -
|\Psi^{\perp}\r |\Psi\r )}{\sqrt{2}}=
\frac{(|0\r |1\r -
|1\r |0\r )}{\sqrt{2}}
\end{eqnarray}
is the singlet state.  Charlie now measures his particle
along the axis corresponding to the states $|0\r$ and 
$|1\r$.  Whatever result he obtains for his particle,
the other two particles will be in an entangled state shared
by Alice and Bob.  For example, if Charlie finds his
particle in the state $|1\rangle$, Alice and Bob
share the state in Eq. (\ref{3.1}).  Note that Charlie
can choose the states with which the state $|\Psi\r$
will be entangled by choosing the axis along which to
measure his particle.

This implies that 
if  we want to produce either the entangled state of 
$|\Psi\rangle$ with
$|0\rangle$ or the entangled state of $|\Psi\rangle$ 
with $|1\rangle$, and we 
don't care which one we get, this can be done with perfect fidelity.
Perhaps a better way of stating this is that if we want to entangle
$|\Psi\rangle$ with one of two orthogonal states, this can be done
perfectly, and we will know with which state it is entangled.

\section{Conclusions}
\label{sec6}
In this paper we have studied various possibilities for entangling
two qubits so the initial information about their preparation is
preserved. We have studied
a specific situation when the state of one of the qubits is known while
the second state is arbitrary. We have shown that entanglement 
via symmetrization in
this case can be performed with a very high fidelity (much higher than
the fidelity of estimation). This type of entanglement 
can be very useful for 
stabilization of  the storage of an 
(unknown) quantum state of one qubit against environmental interaction
and a random imprecision  \cite{Barenco}.
We have shown that the U-NOT gate {\em optimally} 
implements the entanglement transformation 
$|\Psi\r \rightarrow |\Psi\r|\Psi^\perp\r + 
|\Psi^\perp\r|\Psi\r$. This means that the transformation (\ref{4.2}) 
is very special indeed - it describes the optimal cloning, the optimal
U-NOT transformation as well as the optimal entangler.

\acknowledgements
We thank Nicolas Gisin and Christoph Simon for helpful 
 discussions.
This work was supported by the National Science Foundation
under grant PHY-9970507 and by the IST project EQUIP under the contract
IST-1999-11053.

\end{multicols}
\end{document}